\documentclass [12pt]{article}

\begin{document}

\begin{center}
\textbf{The wave function as a mathematical description of a free particle's 
possible motion:}
\end{center}

\begin{center}
\textbf{The 2-slit problem with attempted detection} 
\end{center}

\begin{center}
\textbf{T F O'Malley} 
\end{center}

\begin{center}
\textbf{1746F S Victoria Ave {\#}283, Ventura, CA 93003, USA}
\end{center}

\begin{center}
\textbf{tfomalley@earthlink.net}\textsf{} 
\end{center}

\begin{center}
\textbf{Abstract}\textsf{} 
\end{center}

Starting with a down to earth interpretation of quantum mechanics for a free 
particle, the disappearance and reappearance of interference in the 2 slit 
problem with a detector behind one are treated in detail. A partial 
interpretation of quantum theory is employed which is simple, emphasizing 
description, yet adequate for addressing the present problem.

Given that the eigenvalue equation is essential to predict a free particle's 
probability of collision, it is argued that there is equal need for a 
realistic theory to describe its possible motion. Feynman's point-to-point 
space-time wave packet is put forth and used as the appropriate description 
of the field-free motion between collisions. 

For a particle in a conventional 2-slit experiment with attempted detection 
behind one, the disappearance of interference is explained - both when the 
detection succeeds and when it doesn't. Also a definite prediction is made, 
when the inter-slit distance is reduced, of where the first signs of 
interference should appear on the detection screen.

\newpage

\textbf{1. Introduction} \textsf{} 

The 2-slit experiment, in which a barrier with two slits between a wave's 
source and a detection screen causes an interference pattern, is central to 
many of quantum mechanics' puzzles - measurement, apparent electron 
bi-location, wave collapse and even non-locality.

The quantum interference effect has been studied with photons, electrons and 
heavier particles. With electrons investigations have been done by Jonsson 
[1], by Lichte [2], and Tonomura et al [3], and with atoms by Chapman et al 
[4].

In addition to showing the expected interference in the simple 2-slit 
experiment, the more puzzling result has always been that any attempt to 
answer the question ``which slit did the electron go through'' by attempting 
to detect it behind one or both slits causes the interference to disappear. 
This disappearance of the interference is a phenomenon which obviously needs 
understanding.

In a recent book, Penrose [5] looks at all the most popular attempts that 
are being made to understand this and other quantum puzzles. Although some 
approaches seem more promising than others, all are still referred to as 
works-in-progress with a fuller understanding still to be determined.

The present approach for a free electron, or any quantum entity, interprets 
the time-dependent Schrodinger equation with static potentials as an 
equation of motion and its wave function as a description of an electron's 
possible space-time motion from a starting point to the location of a 

\noindent
possible interaction in its forward direction - ending where and when is 
encounters the interaction. Treating a free particle's wave function as 
simply a description of space-time motion is the way in which a solution of 
Newton's equation, and also the wave function $\Psi $ in quantum field 
theory (n=0), are understood.

We look for a space-time solution, not of the eigenvalue equation (Eq. 1 
below) but of the time-dependent equation (Eq. 2) with one or more static 
potentials V(r$_{i})$, in the electron's forward direction - a solution 
which may consist of one or more wave packets each ending at the site of a 
potential collision. These wave packet solutions are applied to the 2-slit 
problem with an attempted detection behind one of the slits - in order to 
find an explanation for both the absence of interference and its 
reappearance when the inter-slit distance is made sufficiently small.

The approach has much in common with the probabilistic interpretations of 
the Schrodinger wave function in that the wave is not treated as a real 
physical entity but as a description of possible motion, and also in that 
the motion is related to a possible interaction - which is a precondition of 
knowledge. In focusing on a possible localized interaction it is closest to 
Keller's [6] probabilistic proof of collapse, and has something in common 
with that of Ghirardi et al [7] in which a wave may be collapsed by the 
environment. But the collapse in Keller's and the present approach is 
simpler - needing only a single interaction in its environment.

Because it focuses on a free particle with multiple potential interactions 
V(r$_{i})$ in its path it is analogous to many histories theories [8,9,10]. 
Its future histories are also superpositons but ones which describe a 
possible space-time motion toward each of the competing potential 
interactions. 

The approach has even more in common with that of Zurek [11], but with his 
environment for a free electron identified with one interaction among a 
number of possibilities. 

.

It also resembles the deBroglie-Bohn theory of electron motion [12] which 
likewise emphasizes motion along definite paths. However their individual 
paths are single classical paths, rather than Feynman's broad quantum 
superpositions of such paths. .

Finally the approach owes most to Feynman [13]. The point-to-point wave 
packets we recommend to describe space-time motion for a free particle and 
to explain the disappearance of interference were shown by him to be 
solutions of the time-dependent Schrodinger equation, but solutions whose 
description of particle motion between collisions is very different from the 
unrealistic plane plus outgoing wave description in the eigenstate solutions 
of standard quantum theory.

\textbf{2.Setup for the electron 2-slit problem} 

\textbf{2.1 The experiment and its geometry}

It is assumed here that an electron is emitted from a source S and 
ultimately detected at a fluorescent screen D which is distant from the 
source by at least several centimeters (10$^{9}$ atomic units). Between S 
and D is a barrier with 2-slits, A on the left and B on the right, which are 
separated by a lesser macroscopic distance d. In the most interesting 
versions of the experiment, something to monitor or detect the electron's 
presence is inserted just behind a slit (assumed to be slit B here). Since 
experiments have shown that the interference disappears whether or not the 
attempted detection has succeeded, the present objective is to analyze what 
is happening in either case.

\textbf{2.2 The detection method}

It is assumed that the attempt to detect which slit it ``went through'' is 
done in a manner which was described by Feynman [14]. He suggested that we 
try to see the electron by aiming a beam of photons at an area immediately 
behind one of the slits (slit B). The desired result is to bounce a real 
photon off the wave, transferring energy and momentum to the electron wave 
(a Compton effect), and thereby try to verify that it went through B. It is 
further assumed here that the beam is sufficiently dense that it is 
impossible for the electron to go through B and continue without first being 
detected.

This introduces a second possible interaction, the detection, to the problem 
in addition to the necessary detection screen D. (The word ``detection'' 
will be used here to mean ``attempted detection'' whether or not an 
interaction with a photon actually succeeds and regardless of whether or not 
it is directly monitored (by looking for deflection of a photon).

\textbf{3. The Schrodinger equation with static potential}

\textbf{3.1 The eigenvalue equation} 

Schrodinger theory is a generalization of Newton's equation of motion. 
Schrodinger postulated his eigenvalue equation 


\begin{equation}
    \left[ -\frac{\hbar^{2}}{2m}\nabla^{2}+ V(r)\right]\Psi = E\Psi
    \label{eq:1}
\end{equation}

\noindent
to predict accurately the discrete energy levels of the Hydrogen atom with 
the potential term V(r) being the electrostatic potential energy between the 
electron and a massive positive charge. With the inclusion of electron 
exchange it makes similar successful predictions for the permanently bound 
states of all other atoms and molecules. In general it does the same for the 
harmonic oscillator, and for a particle inside a potential well or 
permanently confined inside the (sometimes infinitesimal) range of a 
potential V(r).

Its purely spatial wave function $\Psi $(r) could be considered a fuzzy 
description of the particle's presence everywhere inside the range of the 
potential and is defined independently of the space-time world outside of 
this range. The theory also allows the appending of an external periodic 
time factor which does allow an eigenstate to relate to the outside (e.g. in 
a superposition with other eigenstates or in other ways).

The eigenvalue equation is also used in standard quantum theory for 
non-eigenstates, i.e. for propagating particles. What it does correctly and 
adequately is to predict the electron's collision cross section at the site 
of an actual or possible interaction, and hence its probability of 
occurrence. This means that it \textit{is} correct inside the generally very limited 
range of the potential.

What it does \textit{not} do correctly is to describe the motion of a free electron in 
space and time outside the potential's range, from where it starts to its 
expected destination at the site of the potential. Solutions of Eq. 1 show 
an electron as a plane wave coming in from anywhere and going out from the 
potential in all directions at once. But in reality ``an electron goes from 
place to place'' [15] while it moves field-free from a starting point toward 
the location of a finite range potential interaction V(r). It does \textit{not} move 
from anywhere to everywhere. 

Another failure of Eq. 1 is that it assumes the electron is always inside 
the range of the potential V(r) which is not true for a propagating 
particle. It doesn't allow an electron to be free while it is moving outside 
the range of both the previous interaction left behind and the possible 
interaction it is moving toward. This overlooks the fact that the 
interaction's range is finite and extremely small ($\sim $ 10$^{ - 4}$ cm in 
the present case with visible photon detection, and would be 10$^{ - 8} $ cm 
for atomic collisions). This 10$^{ - 4}$ cm range is infinitesimal compared 
with the distances to and from the slits which are many centimeters. So the 
2-slit photon is completely free on 99.99{\%} of its motion until it finally 
ends at the location of the possible collision.

So while the eigenvalue equation is essential for predicting what is 
possible for a propagating particle (the cross sections), there is need for 
a more general Schrodinger equation (Eq. 2 below) which allows an electron 
to be field-free and to move from place to place - and for a solution of 
that equation (Feynman's) which can describe this motion realistically.

\textbf{3.2. The time-dependent equation of motion for a free propagating 
electron} 

The full time-dependent Schrodinger equation is


\begin{equation}
    \frac{\partial \Psi}{\partial t}= \left(\frac{-i}{\hbar}\right)
    \left[-\frac{\hbar^{2}}{2m}\nabla^{2} + V\left(
    r\right)\right]\Psi \left(\mathbf{r}, t\right)
    \label{eq:2}
\end{equation}

\noindent
where V(r) is a potential interaction of finite, generally very short range 
in a free electron's forward direction.

In the present 2-slit problem with a detector behind one slit or in the 
atmosphere or in gas dynamics studies (where an electron has possible 
interactions with as many as 10$^{20}$ atoms [16]), a free point electron or 
other particle generally confronts more than one possible interaction. For 
that reason the V(r) in Eq. 2 should more properly be written as a 
superposition


\begin{equation}
    V\left( \mathbf{r}\right) =\sum_{i}V\left( \mathbf{r}_{i}\right)
    \label{eq:3}
\end{equation}

An important advantage of Eq. 2 over the eigenvalue equation is that it 
allows intrinsically space-time-dependent solutions [13] rather than purely 
spatial waves with an external time factor. Also its solutions may go 
directly from one place to another as do real particles.

\textbf{3.3 The Feynman wave packet solution of the time-dependent equation}

Feynman defined a space{\-}time wave packet, over the range where $V(r)$ is 
essentially zero, as a superposition of free space{\-}time paths with each 
having its own proper time. (In other words both position and time are 
somewhat uncertain, something detailed further in Sec. 3.4.) From its
starting point, the wave packet's description of the particle's motion 
spreads out symmetrically around the vector ri and then converges to and 
terminates at the local site of a possible interaction [at a V(r)]. He
showed that this construction solved the time dependent Eq. 2 from start to 
end point at the V(r), even though its initial and final boundary conditions
make it incompatible with Eq. 1.

Unlike the corresponding unrealistic solutions of the eigenvalue equation, 
the wave packet's intrinsically time{\-}dependent solution does not extend 
past the site of the finite range potential V(r$_{i})$ to
where the particle
would be in a different energy/momentum state of motion {\-} and where the
outgoing spherical wave solution of Eq. 1 describes unphysical motion.

For the simplest collision problems there is only one potential interaction 
and the solution of Eq. 2 is a single wave packet terminating at that site. 
For the 2-slit problem with attempted detection at \textit{both} slits, the solution of 
Eq.2 would also be a relatively simple superposition of the two wave packets 
terminating at A and B respectively. The present paper concentrates on the 
slightly more complicated case of an attempted detection at only one slit 
(B). 

\textbf{3.4. Space and time uncertainties for the Feynman wave packet 
solution} 

This subsection explores two interesting properties of a point-to point wave 
packet solution of Eq. 2 (as opposed to a general solution of the 
Schrodinger equation), deriving them by using the uncertainty principle. 
Although the uncertainty principle gives only an upper bound it has been 
shown that, for waves whose frequency or wave-number distribution is smooth 
and even roughly approximates Gaussian shape as is the present wave packet 
solution of Eq. 2, the principle is a good approximation. It will be used 
here to explore both the wave packet's space and time uncertainties.

Consider a free wave packet propagating forward over a macroscopic distance 
D in the laboratory from one local position to another, and with an average 
momentum p whose wavelength is much less than D. From the deviations from a 
perfect wave fit between its start and end points one finds approximately 
that the wave packet's uncertainty or spread in momentum in its forward 
direction is 


\begin{equation}
    \Delta p\approx \frac{\hbar}{D}
    \label{eq:4}
\end{equation}¥

>From this, the spatial uncertainty in the forward direction is


\begin{equation}
    \Delta x= \frac{\hbar}{\Delta p}=D
    \label{eq:5}
\end{equation}

The wave packet's uncertainty and its forward extension in space are the 
same. In other words its spatial resolution undefined over approximately its 
entire length and it could as well be anywhere in that span, however long.

The time uncertainty $\Delta $t may also be found, starting from 


\begin{equation}
    \Delta E \approx p\frac{\Delta p}{m}=\sqrt{\frac{2E}{m}
    }\Delta p \approx \hbar \frac{v}{D}
    \label{eq:6}
\end{equation}

\noindent
where (2) has been used, and v is the corresponding classical velocity. From 
this, the time uncertainty is


\begin{equation}
   \Delta t\approx \frac{\hbar}{\Delta E}= \frac{D}{v}
    \label{eq:7}
\end{equation}

This is just the classical time of flight over the entire macroscopic path. 
So, for a wave packet, time (as well as distance) is completely unresolvable 
and therefore meaningless in the center of mass frame of a free point 
electron or other particle. The time uncertainty is analogous to the time 
independence of the photon in its own frame of reference, to that of an 
atom's timeless eigenstate, and it, along with Eq. 5, is also at least 
consistent with the non-locality of propagating quantum particles detected 
by Aspect [17] for photons. It is interesting that a more complete time 
irrelevance is also found at the quantum gravity scale (see for example 
Barbour [18]).

\textbf{4. The wave packets with no attempted detection} 

\textbf{4.1 The one slit wave - $\Psi $}$_{SAD}$ 

Although the present paper is concerned only with the 2-slit problem, and 
there only with attempted detection (Section 5 ff), the purpose of this 
section is to introduce the appropriate Feynman wave packets in the context 
of an experiment with slits, - starting with one slit.

If there is only one open slit A, the wave for possible electron motion 
extends from the source S through slit A and, after deflection, continues to 
the screen D. Even for this simple example, the potentials V(r) must include 
first the turning point at slit A and, from A, all of all the potential 
interaction points on the  detection screen, presumably with individual 
molecules. It may be written as


\begin{equation}
    V_{SAD}=V\left( r_{SA}\right) + 
    \sum_{i}V\left( R_{ADi}\right)
    \label{eq:8}
\end{equation}

\noindent
where the plus sign (+) here is meant to signify ``followed by'', and with 
the r$_{ADi} $being the vector distances from A to the all the active points 
on the screen.

For these potentials the solution of Eq. 2 for electron motion is a wave 
packet from S to A followed by a superposition of minor wave packets from A 
to the screen D, i.e.


\begin{equation}
    \Psi_{SAD}=\Psi_{SA}+\sum_{i}\Psi_{ADi}
    \label{eq:9}
\end{equation}

\noindent
where the plus sign again means ``followed by''. $\Psi _{SA}$ is the wave 
packet from S to A, and the $\Psi _{ADi}$ are all those with paths going 
from A to the screen. $\Psi _{SAD}$ is illustrated in Fig. 1 with several 
wave packets to the screen and a few Feynman paths illustrated for each 
packet.  

\textbf{4.2 The 2-slit wave (with no attempted detection)} 

This is the standard 2-slit problem. With a second slit B, there is added a 
second set of potentials V(r$_{BD})$ for a possible passage of the electron 
through B. With 2 partially different sets of possible interactions (A and 
B) for an electron starting at the source S, the 2 possibilities allow a 
superposition of $\Psi _{SAD}$ with its mirror image $\Psi _{SBD}$ going 
from S to the same points on the screen through slit B

Behind the slits the wave packets through, B with their paths, must cross 
(not shown) and therefore interfere with those through A on their way to the 
same targets on the detection screen D - as always observed. (They may cross 
because the non-crossing rule used in Ref. 12 for trajectories applies only 
to the purely spatial paths in eigenstates and was not proved for space-time 
paths.)

The philosophical question ``which slit?'' for this case is outside the 
scope of the present investigation and is not considered here.

\textbf{5. The 2-slit problem with attempted detection behind slit B - no 
interference - $\Psi $}$_{SD} $

When a detector is put behind B, this changes the situation. Before a 
detection was attempted, all wave packets for possible motion would have 
crossed and, after going through A and B, terminated at the screen. But with 
a detector behind B the situation is different. The leftmost branch $\Psi 
_{SAD}$ through slit A is unchanged. But the wave packet $\Psi _{SD}$ 
which would have gone from S through B and crossed the paths from A now 
faces a possible interaction V(r$_{SB})$ whose small range is of order 
$\lambda $ at that site. The B wave packet therefore terminates at that 
location by the packet's point-to-point definition, as pointed out in 
Section 3.3. This effectively truncates all of its previous paths through B 
at the limits of this relatively mesoscopically small region.

The overall wave function from the source, with a detector, is therefore the 
super-position


\begin{equation}
\Psi_{SD}=\Psi_{SAD} +\Psi_{SB}
    \label{eq:10}
\end{equation}

The plus sign (+) here is meant to signify a superposition of the 2 wave 
packets. This wave is illustrated in Fig. 2 assuming that the inter-slit 
distance d is much larger than the range of the potential interaction 
V(r$_{SB})$ at B where $\Psi _{SB}$ terminates.

[If d is not much smaller than the potential's range however, the truncation 
leaves some paths still inside its interaction range of V(r$_{SB})$at B. 
This situation is illustrated in Fig. 3 and is pursued further in Section 
7.]

With the two major waves in (10) terminating at different locations, the A 
packets at the screen D while the B packet terminates at the detection 
region behind B, the question is not ``which slit did the electron go 
through?'' but rather ``which interaction, that at B or that at D, occurred 
first?'' But either way, whether the attempted detection at B is successful 
or not, there are no interfering Feynman paths between the barrier and the 
screen - and so no possibility of interference - just has always been 
observed.

\textbf{5.1 If the electron is} \textbf{\textit{not}}\textbf{ detected 
behind slit B (null measurement case)}

This case is a simple application of the last paragraph. If nothing is 
detected at B it means that the interaction has been at the screen with one 
of the minor wave packets of $\Psi _{SAD}$ from slit A rather than at B. 
Motion from S to slit B (described by $\Psi _{SB})$ is no longer possible. 
With no B paths interfering with motion from A (Fig. 2), there is no 
possibility of interference in this case as pointed out above.

The case of a successful detection at B is considered in the next section.

\textbf{6. If the electron} \textbf{\textit{is}}\textbf{ detected behind 
slit B}   

If the electron is detected behind B (here by interacting with a photon from 
the beam), its location is fixed momentarily at that interaction site, whose 
size is of order of the electron's wavelength $\lambda $. The motion 
described by $\Psi _{SAD}$ is no longer a possibility, and the electron is 
localized momentarily in the interaction region behind B and ready to move 
on in a new energy-momentum state.

There is nothing mysterious about the location of a particle when it 
actually interacts. It has arrived, and is located, at the place where it 
interacted. This situation looks very much like what is generally ascribed 
to a mysterious ``wave function collapse''. Although an actual interaction 
and collapse are outside the limited range of the wave function describing 
motion of a free particle, interactions do happen in the laboratory (and in 
both classical and quantum field theory) and can't be ignored. The following 
subsection reviews some existing evidence showing that collapse of a wave 
function for particle motion occurs at any interaction which may result, 
directly or indirectly in detection or in human knowledge.

\textbf{6.1 .Collapse} 

Indications of wave function collapse have been seen since the earliest days 
of quantum theory - for photons, electrons and heavier particles. The 
following are a few relatively recent works tending to confirm its reality.

Ghirardi et al [7] have used an assumption of random collapses of a real 
wave to successfully demystify certain quantum paradoxes. (In the present 
interpretation collapse is associated with any detection interaction.)

A recent paper by Keller [6] demonstrated rigorously for a propagating 
particle that its wave function, as an amplitude for location in space , is 
collapsed by any detection. Using conditional probability theory, he proved 
that the probability amplitude wave for the location of the ongoing quantum 
particle immediately following the observation becomes limited to the 
observation area, in other words its effective size is collapsed to that of 
the observation area. 

Gas dynamic studies have for many decades successfully predicted the 
behavior of the entities involved by means of classical models. These models 
treat the entities as particles moving straight from one 
momentum/energy-transfer collision to another, and another.. In retrospect, 
now that we know the particles move quantum-mechanically, the success of 
these models shows that the quantum waves for the motion of these gas 
entities (atoms or whatever) are located (localized, collapsed) at the site 
of each collision or interaction. An example of one such analysis is that 
done by Einstein [19].

Many more recent studies of electron mobilities and diffusion in atomic 
gases by a number of investigators such as Pack an Phelps [20], Huxley and 
Crompton [21] have included in their calculations and predictions, the fact 
that an electron's motion is described by quantum mechanics. In the 
Boltzmann models, they replaced the atom's classical cross section with its 
quantum momentum-transfer cross section as seen by the electron. With this 
model and accurately calculated cross sections they predicted mobility and 
diffusion precisely, while continuing to treat the electron as propagating 
on a vector path from one collision to the next, in other words as behaving 
like a collapsed particle at every energy/momentum transfer.

A very recent paper by Borghesani and O'Malley [16] found new and more 
detailed evidence consistent with collapse from electron mobility 
experiments in Neon. Its density and temperature dependence showed that, 
even immediately \textit{before} each collision, the wave packet for electron motion 
locates the potential interactions within the microscopic area of the 
electron's wavelength $\lambda $ squared.

In the present interpretation ``collapse'' means simply that a propagating 
electron or other quantum particle is present momentarily where and when it 
collides - something which it seems difficult to dispute.

\textbf{6.2 After the interaction and collapse}\textsf{} - 
\textbf{$\Psi $}$_{BD}$

After the collapse then, with the electron localized (collapsed) at the 
interaction region behind slit B and ready to move on in a new different 
energy-momentum state, its only possible interactions are the V(r$_{BDi})$ 
at active points on the detection screen D. Its total potential, as in 
Section 4.1, may be written as a superposition of potential interactions at 
all active points on the screen.


\begin{equation}
    V_{BD}=\sum_{i} V\left( r_{BDi}\right)
    \label{eq:11}
\end{equation}

The corresponding wave packet solution $\Psi _{BD}$ of Eq. 2 for electron 
motion in this multiple potential (essentially the mirror image of that
from A to D in Section 4.1 and Fig. 1) must be a superposition of all minor 
wave packets from B to each of the V(r$_{BDi})$ at D, and may be written as


\begin{equation}
\Psi_{BD}=\sum_{i}\Psi_{BDi}
    \label{eq:12}
\end{equation}

. Again, as in Section 5, there are no competing paths from the other slit 
to the screen for it to interfere with and therefore no interference.

The upshot from Sections 5 and 6 is that the presence of a possible 
detection interaction behind one slit eliminates all interference (assuming 
very large slit separation d) - both when the interaction at the slit 
succeeds and when it doesn't. Its return when d is reduced is explored in 
the next section.

\textbf{7. Reducing the inter-slit distance d from infinity toward $\lambda 
$}  

In order for a detecting photon to distinguish sharply between A and B, the 
A-B inter-slit distance d should be much larger than the photon's resolving 
power (of the order of its wavelength $\lambda )$. Such a value $\lambda $ 
for the distance separating interference from non-interference was confirmed 
recently by Chapman et al [4] using Sodium atoms and a somewhat different 
geometry and procedure. They found that, as the distance d is getting close 
to $\lambda $, the interference tends to return.

If we start with a very large d for which no interference is visible, as d 
is decreased and its size begins to approach that of $\lambda $ from a 
distance, there are paths (shown inside the circle in Fig. 3) which the 
truncation at B does not eliminate because they are inside the range of the 
potential V(r$_{SB})$. Fig 3 represents this situation immediately after the 
detector has been inserted and before an actual interaction occurs, whether 
it is at B or at the screen. With decreasing d these B paths begin to cross 
the nearest paths of possible motion from A to the screen, as shown in the 
figure. The first sign of interference caused by these crossings should 
begin to be visible - with the details depending on whether the detection 
behind D was successful or not.

[Note that Fig. 3, as compared with Fig. 2, has been drawn with a constant d 
and the size of the potential interaction at B greatly exaggerated - because 
it is only the ratio d/$\lambda $ which matters. Also for all the figures it 
should be noted that were not found by solving Eq. 2 directly. Because the 
Feynman wave packets always consist of a straight line vector to the 
destination surrounded by symmetric curved space-time paths, no new 
calculation is needed. Only a few packets to active points on the screen are 
shown schematically for illustration with a very small number of surrounding 
paths for each.]

Crossing paths means that amplitudes are added. A crossed wave path going to 
the screen, after oscillating as it propagates, should result in either 
brighter or dimmer points at the screen, i.e. in part of an interference 
pattern. Interference should be seen whether the interaction fails and the A 
paths go on to interact at the screen, or if it succeeds at B and it is the 
subsequent B paths which ultimately do so.

In the next 2 subsections we investigate what is predicted to be seen in 
each case when d is decreased in this way and interference first becomes 
noticeable.

\textbf{7.1 In the null measurement case (no interaction behind B)} 

The electron, as in Section 5, is confronted in its motion from the source S 
with possible interactions either at B or at the screen as shown in Fig. 3. 
If it does not interact at B this means that it does so through its major 
wave $\Psi _{SAD}$ with paths coming from A. As the figure shows, it is 
the rightmost paths from A which must first cross the truncated paths at B. 
So these crossing paths from A will be the first to carry some interference 
to the far right side of the screen in the figure.

If d/$\lambda $ is decreased further, additional A paths will be crossed by 
more B paths and proceed to the screen with the interference they carry 
spreading toward the center with larger d, and also becoming stronger 
because of the additional B paths crossed. 

.

\textbf{7.2 When the electron does interact behind B} 

The initial state of the wave function before the interaction is again that 
shown in Fig. 3 with short B paths toward the screen truncated inside and 
beginning to cross some B paths also trapped inside the B interaction 
region. Interaction initially is only a possibility.

An actual interaction at B localizes the electron there (inside the circle) 
and eliminates the possibility of motion through A whose possible motion was 
described by the corresponding wave packets from A to the screen.

Starting from the electron wave now collapsed inside of the circle as an 
initial condition, with the amplitudes of some of its paths modified by 
interference from A paths and facing the new possibility of interactions at 
the potentials V(r$_{BDi})$, the post-interaction wave solution ($\Psi 
_{BD})$ of Eq. 2 is illustrated in Fig. 4. It must be a superposition of 
all the minor wave packets proceeding from the B interaction region to the 
screen. For the electron, one of these packets must actually succeed in 
reaching the screen and interacting there with its amplitudes possibly 
modified.

As shown in the preceding subsection and seen if Figs. 3 {\&} 4, the first B 
paths carrying interference are those pointing toward the center of the 
screen. As d decreases further more and more A paths are trapped inside the 
circle and spread the interference to b paths in both directions but more 
rapidly to the left than to the right. Also the number of A paths crossed 
and therefore the intensity of the interference increases rapidly with 
decreasing d. 

. 

\textbf{7.3 The overall experimental result predicted for reduced inter-slit 
distance d}

It is assumed here that the simplest experiment is done, with no attempt at 
monitor0ing deflected photons, but by directly observing only the earliest 
emergence of interference patterns on the screen as the inter-slit distance 
d is decreased. In this case the predicted interference should begin as a 
weighted sum of the two effects described above in Sections 7.1 and 7.2.

With decreasing d (larger circle), the predicted interference on the screen 
should spread out and also become stronger. In the failed detection case the 
interference is predicted to spread out from the far right end of the screen 
(nearest to the detector). With a successful detection at B the interference 
is predicted to spread from the center of the screen and more rapidly to the 
left (away from the detector) than to the right with decreasing d.

One additional thing which such an experiment might show is the relative 
frequency of successful to unsuccessful detections for a particular 
attempted detection method.

\textbf{8. Summary} 

The electron 2-slit problem with the standard geometry and an attempt at 
detecting the electron behind one has been considered with the 
time-dependent Schrodinger equation for a free particle understood as an 
equation of motion, and with its wave function solution interpreted as a 
mathematical description of its possible or actual motion. The possible 
motion for this problem is described by a superposition of Feynman's 
point-to-point space-time wave packet solutions of the equation with each 
wave packet related to a particular potential interaction V(r$_{i})$ 
confronting the electron.

A surprising property of these wave packets was found in the process. They 
apparently have the property that both time and distance are totally 
unresolvable in the free particle's own center of mass reference frame.

The present approach was found to explain the disappearance of interference, 
both for successful and for unsuccessful photon-mediated detection behind 
one slit. The case in which the inter-slit distance is gradually reduced 
from very large was also treated in detail, and definite new predictions 
were made of where the returning interference would first appear on the 
detection screen if only the screen is monitored - both when detection 
behind the slit is successful and when it isn't.

\textbf{Acknowledgments} 

The author gratefully acknowledges generous support from the Contractors 
Register, also valuable help and criticisms from Professor Joseph Keller, 
major encouragement by Professors Larry Spruch and Leonard Rosenberg and by 
John Szitkar, and artwork by Deirdre O'Malley. The work is dedicated to my 
late wife Tanya without whose sacrifice this work would not have been 
possible.

\textbf{References} 

[1] Jonsson C 1961 A Phys 161 454.

[2] Lichte H 1986 Ann Ny Acad Sci 480 175

[3] Tonomura A, Endo J, Matsuda T, Kawasaki T, {\&} Ezawa J 1989 Am. J. 
Phys. 57 117
[4] Chapman M, Hammond T, Lenef A, Schmiedmayer Y , Rubenstein R ,

Smith E and Prichard D 1995 Phys. Rev. Lett. 75 3783{\-}87 

[5] Penrose R 1997 The Large, the Small and the Human Mind (Cambridge Univ. 
Press,

New York) 60 

[6] Keller J B 1990 Am. J. Phys. 58 768.

[7] Ghirardi G C, Rimini A and Weber T 1986 Phys Rev. D 34 470 

[8] Griffiths R B 1984 J stat Phys 36 219 

[9] Omnes R 1988 J Stat Phys 53 957

[10] Gell-Mann M 1993 Phys Rev D 47 3345 

[11] Zurek W H 1982 Phys Rev D 126 1862 

[12] Holland P 1993 The Quantum Theory of Motion (Cambridge University 
Press, Cambridge)

[13] Feynman R P 1948 Rev Mod Phys 20 367

[14] Feynman R P 1965 The Feynman Lectures on Physics (Addison{\-}Wesley, 
Reading MA) Vol. 3 

[15] Feynman R P, QED the strange theory of light and matter (Princeton 
University Press, Princeton NJ 1985) 84-85

[16] Borghesani A and O'Malley T 2003 J Chem Phys 118 2760 

[17] Aspect A, Dalibard Y and Roger G 1982 Phys. Rev. Lett. 49 1804. 

[18] J Barbour J 1999 The End of Time (Oxford University Press, New York)

[19] Einstein A 1905 Annalen der Physik 322 549.

[20] Pack J L and Phelps A V 1961 Phys Rev 121 789.

[21] Huxley L G H and Crompton R W 1974 The Diffusion and Drift of Electrons 
in Gases (Wiley{\-}Interscience, New York)

\textbf{Figure Captions} 

Fig. 1. For the 1 slit problem, the figure depicts the solution of the 
time-dependent Schrodinger equation (1) for possible electron motion from 
source S to detection screen D. Each wave packet going to or coming from the 
slit A is a bundle of Feynman paths from start to destination.

Fig. 2. For the 2-slit problem with an attempted detection behind slit B, 
the figure represents the solution of Eq. 2 when the interslit distance d is 
very large. The major wave through B finds its termination there because of 
the potential detection interaction at that point.

Fig. 3. The same as Fig. 1 but with the reduced inter-slit distance d 
beginning to approach the size of the interaction region. It shows the 
beginning of crossing (interfering) Feynman paths.

Fig. 4. The wave $\Psi _{BD}$ with reduced d after a photon interaction at 
B. With the electron starting anew from B, motion along the paths of the 
major wave $\Psi _{SAD}$ through A is no longer possible. The 
closeness of the slits has also fixed the beginnings of some interfering 
paths from A inside the interaction site. The only remaining sites of 
potential interactions are at the screen as shown.

\end{document}